\documentclass[11pt]{article}

\usepackage[preprint]{acl}

\usepackage{times}
\usepackage{latexsym}

\usepackage[T1]{fontenc}

\usepackage[utf8]{inputenc}

\usepackage{microtype}

\usepackage{inconsolata}

\usepackage{graphicx}
\usepackage{caption}
\usepackage{cuted}

\usepackage{booktabs} 
\usepackage{multirow}
\usepackage{makecell}
\usepackage{xcolor}

\title{LongCLI-Bench: A Preliminary Benchmark and Study for Long-horizon Agentic Programming in Command‑Line Interfaces}

\author{
Yukang Feng $^{1,2,3*}$\quad 
Jianwen Sun $^{1,2,3*}$\quad 
Zelai Yang $^{1*}$\quad 
Jiaxin Ai $^{2,4}$\quad 
Chuanhao Li $^{4}$\\
\textbf{Zizhen Li} $^{1,2,3}$ \quad 
\textbf{Fanrui Zhang} $^{2}$ \quad
\textbf{Kang He} $^{2}$\quad 
\textbf{Rui Ma} $^{5}$\quad 
\textbf{Jifan Lin} $^{5}$\quad 
\textbf{Jie Sun} $^{2}$\quad \\ 
\textbf{Yang Xiao} $^{6}$\quad 
\textbf{Sizhuo Zhou} $^{2}$\quad 
\textbf{Wenxiao Wu} $^{2}$\quad 
\textbf{Yiming Liu} $^{2}$\quad \\
\textbf{Pengfei Liu} $^{2,5}$\quad 
\textbf{Yu Qiao} $^{2,4}$\quad 
\textbf{Shenglin Zhang} $^{1}$\quad 
\textbf{Kaipeng Zhang} $^{2,3\dagger}$ \\
[2mm]
$^1$ NKU\quad
$^2$ SII\quad
$^3$ Shanda AI Research Tokyo\quad
$^4$ Shanghai AI Laboratory\quad
$^5$ SJTU\quad
$^6$ PolyU \\
[1mm]
Project Page: https://github.com/finyorko/longcli-bench
}

\begin{document}
\maketitle

\begin{strip}
  \centering
  \includegraphics[width=\textwidth]{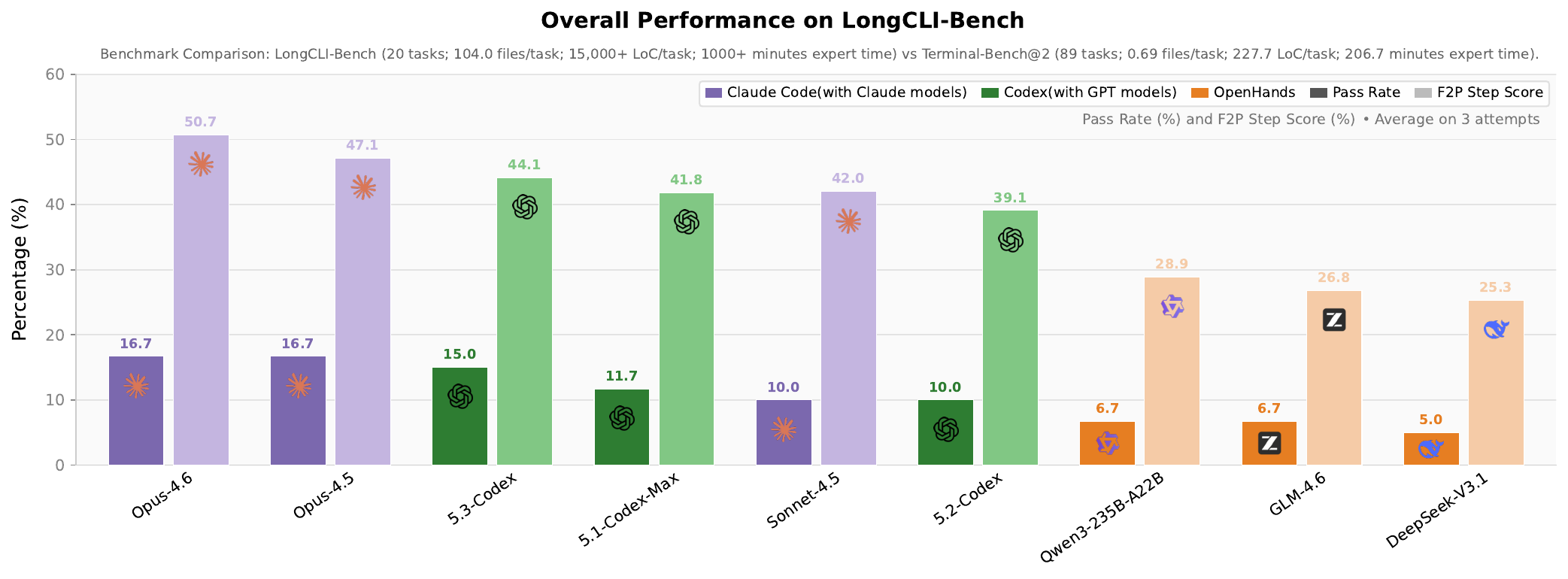}
  \label{fig:overview_performance}
  \vspace{-1em}
\end{strip}

\begin{abstract}
Recent advances in AI-assisted programming have empowered agents to execute complex workflows via command-line interfaces, however, existing benchmarks are limited by short task horizons, data contamination from GitHub scraping, and a lack of fine-grained evaluation metrics, fail to rigorously evaluate the long-horizon planning and execution capabilities essential for realistic software engineering.
To address these gaps, we introduce \textbf{LongCLI-Bench}, a comprehensive benchmark designed to evaluate agentic capabilities across \textbf{long-horizon}, realistic tasks.
We curated 20 high-quality, long-horizon tasks from over 1,000 computer science assignments and real-world workflows, covering four engineering categories: from scratch, feature addition, bug fixing, and refactoring.
We propose a dual-set testing protocol for LongCLI-Bench, which measures requirement fulfillment \textit{(fail\(\to\)pass)} and regression avoidance \textit{(pass\(\to\)pass)}, and incorporates step-level scoring to pinpoint execution failures.
Extensive experiments reveal that even state-of-the-art agents achieve pass rates below 20\% in LongCLI-Bench.
Step-level analysis further indicates that the majority of tasks stall at less than 30\% completion, highlighting that critical failures often occur in the early stages.
Although self-correction offers marginal gains, human-agent collaboration through plan injection and interactive guidance yields significantly higher improvements.
These results highlight that future research must emphasize the development of synergistic human-agent workflows alongside advances in agents’ planning and execution capabilities to overcome key challenges in long-horizon task performance.
\end{abstract}

\section{Introduction}

The paradigm of AI-assisted programming is undergoing a fundamental transition from code generation to autonomous software engineering.
Modern agents such as SWE-agent~\citep{swe-agent}, OpenHands~\citep{openhands}, and commercial CLI assistants~\citep{codex,claude-code,gemini-cli} can now function like human engineers: they plan architectures, navigate repositories, manage development environments, and execute multi-step workflows via Command Line Interfaces (CLI) based on environment feedback~\citep{react,reflexion,self-refine}.

\begin{figure}[t]
  \includegraphics[width=\columnwidth]{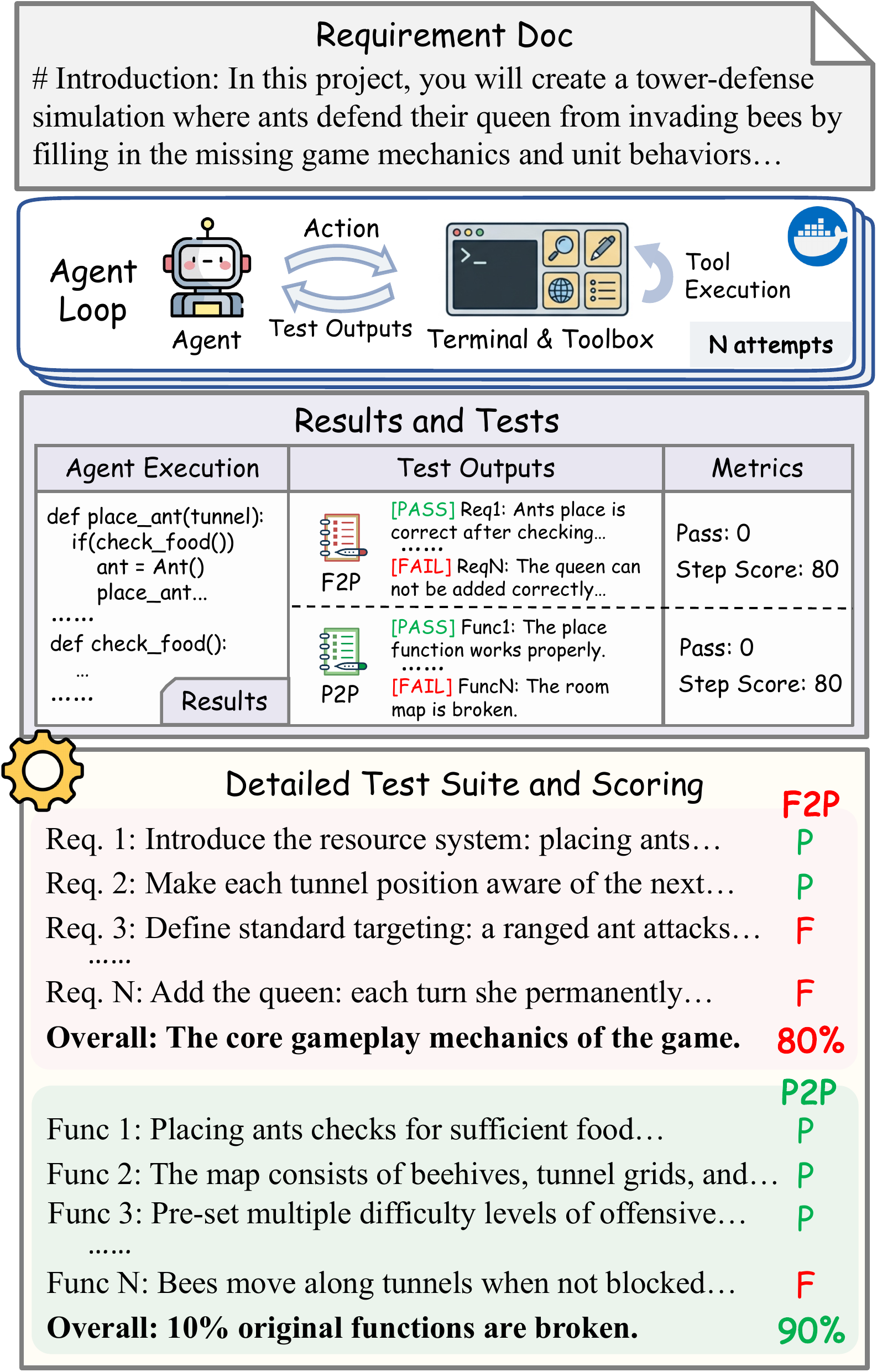}
  \caption{Task sample in LongCLI-Bench.}
  \label{fig:overview}
  \vspace{-1.5em}
\end{figure}

Despite the advancements, evaluation benchmarks have lagged behind.
Early benchmarks, such as HumanEval and MBPP~\citep{humaneval,mbpp}, assess only algorithmic logic within isolated code snippets.
While subsequent benchmarks like SWE-bench~\citep{swe-bench,swe-bench-verified} extend to repository-level, they remain restricted to short, single-category tasks, and are compromised by data contamination from scraped GitHub repositories.
Recent Terminal-Bench ~\citep{terminal-bench} introduces interactive sandbox environments with community-contributed tasks, yet its tasks remain simple with short horizons, and lack fine-grained evaluation to deliver effective feedback or diagnose failure modes.

In real-world scenarios, software engineering tasks are inherently long-horizon with continuous requirements, yet existing benchmarks ignore such sequential dependencies, failing to evaluate the ability to maintain long-term environmental context and navigate complex, interdependent workflows.

To address the above limitations, we introduce \textbf{LongCLI-Bench}, a comprehensive benchmark designed to evaluate \textbf{long-horizon} agentic capabilities in realistic CLI environments, covering diverse task categories and providing fine-grained evaluation. 
Unlike benchmarks that focus on a single task type, LongCLI-Bench spans four engineering categories: \textbf{From Scratch ($0 \to 1$)} which involves building projects from the ground up, \textbf{Feature Addition ($N \to N+1$)} which focuses on enhancing existing codebases, \textbf{Bug Fix ($No \to Yes$)} which addresses diagnosing and resolving bugs, and \textbf{Refactor ($A \to A’$)} which involves optimizing or restructuring code.
We curate tasks from two distinct sources: 958 computer science (CS) assignments and 50 real-world workflows. This approach not only mitigates the data contamination risk inherent in GitHub-derived datasets but also ensures that the benchmark encompasses realistic, complex, long-horizon tasks that align with genuine development scenarios.
To identify tasks that effectively probe the functional boundaries of agents, we manually crafted test inputs to execute these samples via Codex~\cite{codex}, followed by a manual inspection of the execution results.
We found that existing agents already exhibit high proficiency in handling the majority of routine assignments, so we eliminated these tasks and focused on the remaining complex tasks.
For each task, we manually created the requirement documentation, isolated environments, and test suites and finally filtered the pool into 20 high-quality tasks.

Each task includes an initial repository and a requirement document in an isolated environment. 
The evaluation utilizes a dual-set protocol: \emph{Fail$\to$Pass} (F2P) tests verify that the agent has successfully implemented the new requirements, while \emph{Pass$\to$Pass} (P2P) tests ensure that the agent's modifications do not break existing system functionalities (detecting regressions).
Moreover, we provide \emph{step-level scores}, enabling fine-grained measurement of partial progress and pinpointing exactly where long workflows break.

The experiment results show that long-horizon CLI work remains far from solved: all agent systems achieve $<20\%$ pass rate (e.g., Claude Code~\citep{claude-code} with Claude-Opus-4.6 and Codex with GPT-5.3-Codex), while step-level scores reveal that failures are predominantly concentrated in the early stages of tasks.
Furthermore, although self-correction improves performance by leveraging error feedback, it still lags behind methods involving human plan injection or human-agent interaction. This finding strongly indicates that strategic planning and execution proficiency remain the key bottlenecks for current autonomous agents.

\noindent Our contributions are summarized as follows:
\noindent\textbf{LongCLI-Bench}: A curated benchmark of 20 long-horizon tasks, filtered from $>$1,000 samples via manual curation and rigorous validation process to ensure sufficient long-horizon and complex, covering four distinct engineering categories.

\noindent\textbf{Dual-Set Tests with Step-Level Scores}: Assesses both the requirement implementation (F2P) and the regressions (P2P). Additionally, we introduce step-level metrics to measure the degree of task completion, enabling a clear distinction between early-stage failures and near-success outcomes.

\noindent\textbf{Experiment Results}: All agents yield pass rates below 20\%, identifying planning and execution proficiency as the key limitations hindering autonomous software engineering performance.

Rather than focusing solely on boosting agents' standalone planning and execution capacities, future research should prioritize core engineering proficiencies, long-horizon contextual consistency maintenance, and effective human collaboration to navigate complex engineering tasks.

\section{Related Work}

\subsection{Agents and Coding-Oriented LLMs}

Agentic coding performance is shaped jointly by the LLMs and the agent scaffold. 
Firstly, code-oriented LLMs ~\citep{code-llama,starcoder,seed-coder,opencoder} such as DeepSeek-Coder~\citep{deepseek-coder} and Qwen3-Coder~\citep{qwen3} have demonstrated considerable capabilities in code understanding and generation. 
Secondly, SWE-agent~\citep{swe-agent} can interact with real repositories by interfaces, while OpenHands~\citep{openhands} provides an end-to-end framework; other systems explore different avenues, including structured localization and repair ~\citep{autocoderover}, multi-agent optimization~\citep{agentcoder}, planning coding ~\citep{codeplan}, baselines~\citep{agentless}, and retrieval-augmented completion ~\citep{repocoder}.
In practice, commercial CLI assistants~\citep{codex,claude-code,gemini-cli}, such as Codex, integrate innovative paradigms including general tool-use and self-refinement mechanisms~\citep{react,toolformer,reflexion,self-refine}, demonstrating strong intelligent assistance capabilities.

\subsection{Coding/SWE Benchmarks}

The evaluation of code generation has evolved from simple function-level tasks to increasingly complex, environment-grounded repository-level challenges. 
Early benchmarks measured function-level~\citep{humaneval,mbpp,apps} and class-level~\citep{classeval} tasks, which are useful for assessing local correctness but largely overlook cross-file planning.
Consequently, benchmarks expanded to more diverse suites and domains~\citep{codexglue,ds-1000,multipl-e,bigcodebench}, freshness-oriented evaluation~\citep{livecodebench}, and tasks requiring explicit cross-file reasoning~\citep{repobench,crosscodeeval}.

At the repository level, the SWE-bench series~\citep{swe-bench,swe-bench-verified,multi-swe-bench,swe-bench-multimodal,utboost,swe-smith} evaluates real github issue resolution via patches that satisfy project tests.
Complementary benchmarks extend to feature and long-horizon tasks~\citep{fea-bench,gittaskbench,swe-bench-pro}, repository generation~\citep{nl2repo-bench}, and test generation ~\citep{swt-bench}.
Meanwhile, environment-centric benchmarks~\citep{csr-bench,multi-docker-eval,mle-bench} emphasize reproducible execution.
But these benchmarks are mined from GitHub and are susceptible to contamination~\citep{contamination-survey,contamination-naacl}.
Terminal-Bench~\citep{terminal-bench} provides a standardized sandbox with community-contributed tasks for evaluating agents' coding and terminal capabilities. 
However, it is limited to short-duration tasks and offers only binary pass/fail feedback, which prevents a detailed analysis of the agent’s performance.
Furthermore, precisely because these tasks lack the complexity and continuity inherent to real-world workflows, existing benchmarks are increasingly facing saturation.
We therefore introduce LongCLI-Bench, a benchmark of multi-category, long-horizon real-world tasks with step-level evaluation.

\section{LongCLI-Bench}

LongCLI-Bench is designed around five principles: long-horizon, contamination control, clear requirements, solvability, and isolated environments.

\subsection{Data Construction}

\begin{figure*}[t]
  \centering
  \includegraphics[width=0.99\textwidth]{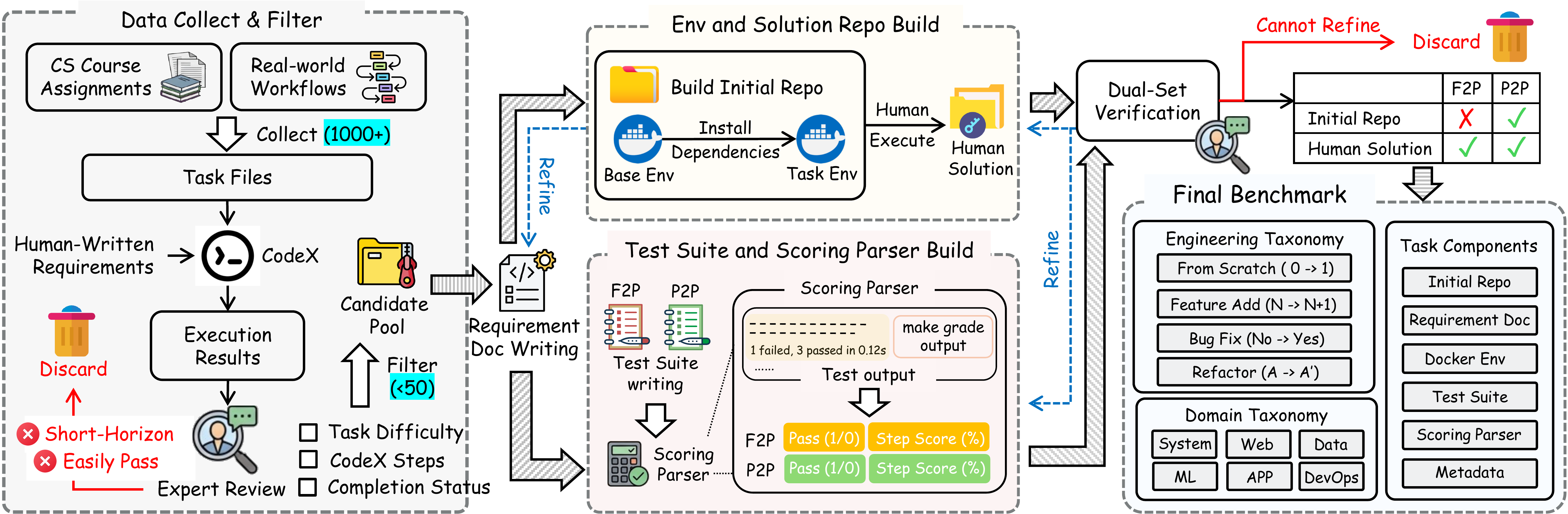}
  \caption{The LongCLI-Bench construction pipeline. We curate tasks from diverse sources and employ a parallel construction method for solutions and tests. The pipeline features a strict Dual-Set Verification mechanism with iterative refinement loops to ensure high-quality, contamination-free benchmarks across various engineering and domain categories.}
  \label{fig:data_pipeline}
  \vspace{-1em}
\end{figure*}

\subsubsection{Data Sources and Initial Codebase}
Most existing benchmarks scrape tasks from popular GitHub repositories, but this increases contamination risk. We avoid direct mining and instead curate tasks from two primary sources: CS course assignments and real-world workflows.

\noindent
\textbf{CS course assignments} are rigorously designed by domain experts and typically feature clear specifications alongside realistic codebases.
However, establishing a benchmark that genuinely challenges modern agents requires extensive filtration. 
To ensure high task quality and alignment with our benchmark goals, we undertook a labor-intensive manual curation process. Specifically, we first identified and collected \textbf{958} assignments from \textbf{108} courses spanning diverse domains such as operating systems, networks, and data processing. 
For each assignment, we organized the task documentation and executed it using Codex. Since most assignments lack automated evaluation metrics, we then manually assessed the solutions generated by Codex to verify task completion status, which allowed us to evaluate the feasibility and difficulty of each task.
Through thorough manual review, we found that Codex excels at most CS homework and many project-level tasks. As a result, we eliminated tasks that were either too easy or hard to evaluate consistently and only a small subset of tasks that required complex, multi-step engineering skills were kept.

\noindent\textbf{Real-world Research \& Engineering Workflows.} These are long-chain tasks derived from actual research and work scenarios, where all information is manually constructed. Inspired by daily development activities, such as writing code, configuring environments, and building data processing pipelines, these tasks represent a sequence of complex, continuous sub-tasks that mirror realistic workflows. Crucially, they exhibit strong sequential dependency, forming a complete requirement chain where the outcome of previous steps determines the feasibility of subsequent ones.

\subsubsection{Requirement Document}
Based on the specific task objectives, we craft a requirement document that explicitly defines both the functional goals and the entry point specifications. This document must clearly articulate the functionality to be implemented.
This avoids false negatives where correct logic fails testing due to arbitrary, unlocatable entry points.

For CS course assignments that come with existing descriptions, we partially rewrite the requirements by replacing specific variable/function/file names, and background stories to prevent simple retrieval matching, thereby reducing contamination risks while enhancing readability. For tasks derived from real-world scenarios, the requirements are entirely manually crafted.

\subsubsection{Environment and Solution Codebase}
We concurrently construct the execution environment and solution codebase within Docker container.
This process involves iteratively solving the task from a base docker image while recording necessary dependencies. The final outputs are rigorously separated: dependencies are solidified into a Dockerfile to provide a consistent base environment, while the human solution path form the solution repo to verify solvability. 

\subsubsection{Test Suite and Scoring Design}
Test scripts are written based solely on the \textbf{requirement document}, not the solution repo. This prevents ``baking in'' the implementation into the tests and avoids inheriting potential errors from the human solution.
We employ two test sets: 1) \emph{Fail $\to$ Pass} (F2P): 
These tests evaluate the completion status of the requirements. 
2) \emph{Pass $\to$ Pass} (P2P):
Often ignored in other benchmarks, these tests verify that the agent's modifications have not broken existing system functionality.

To accurately reflect task completion, we prioritize environment-state verification, such as confirming a service deployment by checking if the port is listening and the log content, rather than just validating that a start command was written. 
Furthermore, both test sets contribute to a \textbf{step-level score}. For CS assignments with built-in grading scripts, we parse the output to calculate this score; for tasks without predefined tests, we manually segment the requirements into sub-tasks and write corresponding evaluation scripts.

\subsubsection{Verification and Quality Control}

\textbf{Verification condition.} A task is valid if the F2P tests fail on the initial repo and pass on the solution repo, while the P2P tests pass on both.

To ensure benchmark quality, we adopt a rigorous iterative, closed-loop verification procedure: 
1) During the environment setup and solution creation, we immediately revise the documentation if any ambiguities or missing content are identified.
2) Test Suite Validation. Execute test scripts against both the initial and solution repos. If the verification condition is not met, it indicates an issue with the test script, the solution, or the environment. Human experts review and fix the issue, triggering a re-validation. If a task cannot meet the verification condition after three iterations, it is discarded.
3) Expert review. Each task undergoes an final expert review to ensure logical correctness and feasibility before inclusion.

\subsection{Task Composition and Evaluation}
\subsubsection{Task Composition}

LongCLI-Bench is designed around five principles: long-horizon, isolated environments, clear requirements, solvability, and contamination control. Each task consists of the following components:

\noindent\textbf{Initial Repo}: The task starting codebase (e.g., a skeleton or empty directory).

\noindent\textbf{Task Requirement}: A task requirement document that explicitly defines functional goals and entry points to ensure clear requirements.

\noindent\textbf{Environment}: An isolated environment ensuring consistent execution and reproducibility.

\noindent\textbf{Solution Repo}: A human-authored ``golden solution'' ensuring solvability and validation.

\noindent\textbf{Tests}: A dual-set testing suite verifying both requirement fulfillment and regression testing.

\noindent\textbf{Scoring Parser}: A parser that analyzes test outputs to calculate a step-level metric.

\noindent\textbf{Metadata}: Includes task type, domain, difficulty level, and estimated human completion time.

\subsubsection{Evaluation Workflow}

Based on above components, the evaluation follows the sequence:

\noindent\textbf{1) Initialization}: 
An isolated docker environment is initialized with the initial repo.

\noindent\textbf{2) Execution}: 
The agent receives the requirement, plans and interacts with the terminal to execute.

\noindent\textbf{3) Test}: 
Upon task completion or timeout, the harness executes the test scripts.

\noindent\textbf{4) Scoring}: 
The Scoring Parser aggregates the test results to compute the pass rate and step scores.

LongCLI-Bench also supports optional evaluation approaches: 

\noindent\textbf{1) Multiple Attempts}: The agent executes the task repeatedly for N independent attempts. 

\noindent\textbf{2) Self-Correction}: The agent leverages test feedback from the prior turn and re-executes to refine the solution in a multi-turn process.

\subsection{Taxonomy}
\subsubsection{Engineering Taxonomy}
To systematically evaluate agent capabilities across the software development lifecycle, we classify tasks into four distinct engineering categories. 
This classification enables targeted assessment of agent performance across a full spectrum of core engineering competencies.

\noindent\textbf{From Scratch ($0 \to 1$)}: The ability to plan, configure, and build a runnable project from scratch.

\noindent\textbf{Feature Addition ($N \to N+1$)}: The ability to add new modules to an existing codebase.

\noindent\textbf{Bug Fix ($No \to Yes$)}: The ability to diagnose, locate, and fix complex bugs.

\noindent\textbf{Refactor ($A \to A'$ )}: The ability to optimize or restructure code.

\subsubsection{Domain Taxonomy}

Beyond engineering types, LongCLI-Bench covers six primary domains to assess agent capabilities across diverse technical fields. 
  
System Programming: Operating system, compilers, memory, concurrency, embedded programming, hardware interfacing, and distributed system.

Web Development: Frontend/backend, database, API design, authentication, and service.

Data Engineering: Crawling, filtering, formatting and statistical analysis/visualization.

Machine Learning: training, inference, deployment, evaluation, and signal processing.

Applications: Business or gameplay, simulation engines, physics integration, and interactive tool.

DevOps: CI/CD pipelines, containerization, environment build, system monitoring/diagnostics.

\subsection{Statistics of LongCLI-Bench}

\begin{table}[h]
  \centering
  \caption{Comparison of LongCLI-Bench and Terminal-Bench@2. Files, LoC and Expert Time are average results; LoC stands for Lines of Code.}
\resizebox{\linewidth}{!}{
  \begin{tabular}{lccccc}
    \toprule
    Benchmark          & Tasks & Files & LoC   & Expert Time (min) \\
    \midrule
    Terminal-Bench@2  & 89    & 0.69      & 227.7     & 206.7       \\
    LongCLI-Bench      & 20    & 104.0     & 15,000+   & 1000+        \\
    \bottomrule
  \end{tabular}
  }
  \label{tab:benchmark_comparison}
\end{table}

We present a detailed statistical analysis of the 20 curated tasks in LongCLI-Bench.

LongCLI-Bench averages 15,000+ lines of code (LoC) and 104 source files per task, covering a diverse linguistic landscape, including C, Python, Java, and JavaScript, representing complex, repository-level engineering.
In stark contrast, Terminal-Bench@2 averages only 227.7 LoC and 0.69 files.
This disparity highlights that LongCLI-Bench evaluates the ability to handle long-horizon complex problems within massive, interdependent systems rather than executing isolated snippets.

This complexity extends to the temporal dimension, imposing a significantly higher cognitive load. Expert completion time for LongCLI-Bench averages 1000+ minutes (vs. 206.7 min for Terminal-Bench@2). These metrics confirm that the benchmark effectively probes long-horizon planning and context maintenance capabilities, differentiating it from short-term execution tasks.

\section{Experiments}

\begin{table*}[t]
  \centering
  \caption{Overall Performance on LongCLI-Bench (Average on 3 attempts). We report the Pass Rate(\%), Pass@3(\%), and average Step Scores(\%) for both F2P and P2P tests. Time is measured in minutes.}
  \label{tab:main_results}
  \resizebox{0.98\linewidth}{!}{
  \begin{tabular}{l l c c c c c c c}
    \toprule
    Agent & Model & Pass & Pass@3 & F2P Pass & F2P Step Score & P2P Pass & P2P Step Score & Time(min) \\
    \midrule
    Codex 
    & GPT-5.1-Codex-Max & 11.7 & 15.0 & 16.7 & 41.8 & 70.0 & 99.2 & 12.9 \\
    & GPT-5.2-Codex     & 10.0 & 15.0 & 13.3 & 39.1 & 73.3 & 99.4 & 10.9 \\
    & GPT-5.3-Codex     & 15.0 & 20.0 & 18.3 & 44.1 & 86.7 & 99.5 & 8.8 \\
    \midrule
    Claude Code 
    & Claude-Sonnet-4.5 & 10.0 & 10.0 & 13.3 & 42.0 & 70.0 & 98.4 & 10.6 \\
    & Claude-Opus-4.5   & 16.7 & 20.0 & 20.0 & 47.1 & 71.7 & 98.7 & 11.3 \\
    & Claude-Opus-4.6   & \textbf{16.7} & \textbf{25.0} & \textbf{20.0} & \textbf{50.7} & 78.3 & 99.1 & 17.6 \\
    \midrule 
    OpenHands 
    & DeepSeek-V3.1     & 5.0 & 10.0 & 11.7 & 25.3 & \textbf{88.3} & \textbf{99.7} & 17.9 \\ 
    & GLM-4.6           & 6.7 & 10.0 & 11.7 & 26.8 & 83.3 & 99.5 & 25.3 \\
    & Qwen3-235B-A22B         & 6.7 & 10.0 & 13.3 & 28.9 & 81.7 & 99.3 & 30.2 \\
    \bottomrule
  \end{tabular}
  }
\end{table*}

\subsection{Experiment Setup}

\noindent\textbf{Agents and Models.} We evaluate two distinct categories of agents.
The first comprises commercial CLI assistants driven by proprietary models: Codex (gpt-5.x-codex series) and Claude Code (claude-sonnet/opus-4.x series).
The second leverages the OpenHands\cite{openhands} framework to assess the leading open-source models, including DeepSeek-V3.1\cite{liu2024deepseek}, GLM-4.6\cite{glm46_blog}, and Qwen3-235B-A22B\cite{qwen3}.

\noindent\textbf{Evaluation Metrics and Protocol.} To ensure fairness, we employed a unified system prompt and averaged results over \textbf{three} independent runs. 
We first measure \textbf{step scores} for both requirement fulfillment (F2P) and regression testing (P2P), assessing the percentage of sub-steps completed. 
A binary ``pass'' is defined strictly: a test set is considered passed only when its step score reaches 100\%. 
Accordingly, we derive the \textbf{F2P pass rate} and \textbf{P2P pass rate}. 
The overall \textbf{Pass Rate} is defined as the percentage of tasks where the agent pass on both F2P and P2P tests.
Additionally, we report \textbf{Pass@3} to assess stability, along with execution time for efficiency analysis.

\subsection{Main Results}

Table~\ref{tab:main_results} presents the overall performance of various models on LongCLI-Bench in a single-turn setting.
The experimental results reveal that LongCLI-Bench poses an extreme challenge to current state-of-the-art agent systems. The pass rate for most models falls below 15\%, with even the top-performing Claude-Opus-4.6 achieving only 16.7\%.

This result compellingly demonstrates that while existing agent technologies excel in short code generation, a significant capability gap remains when facing real-world long-horizon tasks requiring long-term memory, environmental perception, and complex logical planning. Commercial systems significantly outperformed the open-source framework in requirements completion (F2P step score), indicating that general frameworks struggle to adapt to such complex engineering tasks without targeted optimizations.

While the average P2P step scores are very high (>98\%) across all models, the P2P pass rate is notably lower, ranging from 70.0\% to 88.3\%.
The low P2P pass rate indicates that in the tasks where agents did attempt complex edits, they introduced broken modification in nearly 12\%-30\% of cases. This corroborates that agents often lose sight of the broader context when focusing on new features, or struggle to adhere to the instructions as task difficulty improves, resulting in unintended breakage of existing functionalities.

Interestingly, OpenHands with DeepSeek-V3.1 achieved the highest P2P pass rate (88.3\%), likely because its lower execution capability (F2P Score 25.3) limited the scope of its modifications, reducing the surface area for potential errors.

\subsection{Step-level Analysis}

To investigate when models fail to complete tasks, we analyzed the distribution of F2P step score.

Table~\ref{tab:f2p_step_score_dist} reveals that failures are not evenly distributed and the majority of tasks fail at early stages. Across all evaluated agents, the highest percentage of outcomes is concentrated in the <30\% range. This implies that agents encountered severe difficulties at the very onset of the tasks.
Notably, commercial agent-model pairs exhibit a lower proportion of scores falling into low range compared to OpenHands, an indicator of their superior early-stage planning capabilities. This strength helps them navigate early task obstacles through more robust upfront preparation and reasoning.

The few cases in the $[80, 100)$ range highlights the sequential dependency of long-horizon tasks, where a failure in a preceding step directly blocks the subsequent steps.
This distribution highlights step-level scoring’s core value: unlike binary metrics that treat all failures equally, step scores reveal the exact breakage point in the execution chain. This granularity allows us to discern partial progress and diagnose whether agents fail at fundamental planning or specific logic implementation, confirming that LongCLI-Bench effectively evaluates the robustness of the entire workflow.

\begin{table}[tbp]
\centering
\caption{Distribution of F2P step scores. Columns show the percentage of tasks falling into specific score ranges.}
\label{tab:f2p_step_score_dist}
\resizebox{\linewidth}{!}{
\setlength{\tabcolsep}{2pt}
\begin{tabular}{llccccc}
\toprule
Agent & Model & [0,30) & [30,60) & [60,80) & [80,100) & [100] \\
\midrule
\multirow{3}{*}{Codex} 
  & GPT-5.1-Codex-Max & 45.0 & 16.7 & 13.3 & 8.3  & 16.7 \\
  & GPT-5.2-Codex     & 51.7 & 15.0 & 10.0 & 10.0 & 13.3 \\
  & GPT-5.3-Codex     & 41.7 & 13.3 & 15.0 & 11.7 & 18.3 \\
\midrule
\multirow{3}{*}{Claude Code} 
  & Claude-Sonnet-4.5 & 46.7 & 13.3 & 11.7 & 15.0 & 13.3 \\
  & Claude-Opus-4.5   & 40.0 & 23.3 & 6.7  & 10.0 & 20.0 \\
  & Claude-Opus-4.6   & 38.3 & 20.0 & 10.0 & 11.7 & 20.0 \\
\midrule
\multirow{3}{*}{OpenHands} 
  & DeepSeek-V3.1     & 65.0 & 13.3 & 5.0  & 5.0  & 11.7 \\
  & GLM-4.6           & 63.3 & 15.0 & 3.3  & 6.7  & 11.7 \\
  & Qwen3-235B-A22B   & 58.3 & 13.3 & 8.3  & 6.7  & 13.3 \\
\bottomrule
\end{tabular}
}
\end{table}

\begin{figure*}[t]
  \centering
  \includegraphics[width=0.98\textwidth]{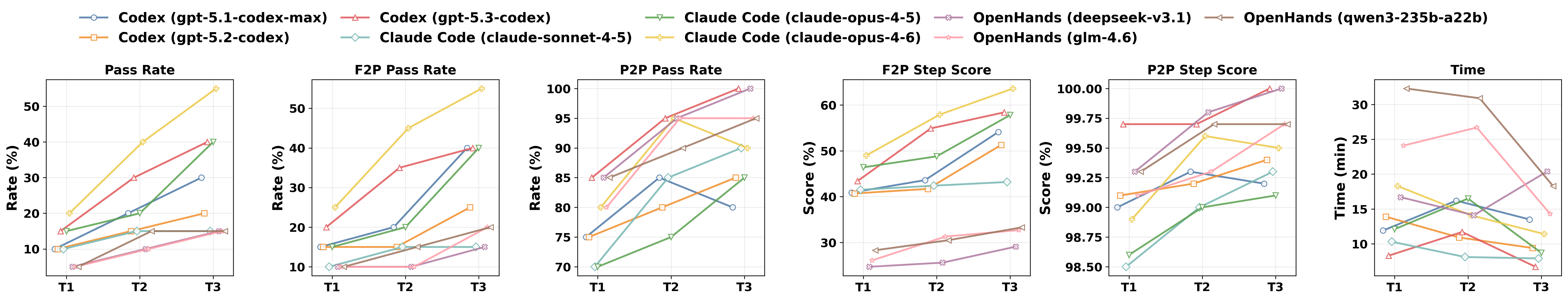}
  \caption{Multi-Turn Self-Correction Performance.}
  \label{fig:selfcorr_turn_metrics}
\end{figure*}

\subsection{Self-Correction Capabilities}

Under the Self-Correction setting, agents re-executes the task using feedback from previous turn enabling multi-turn self-correction.
As illustrated in Figure \ref{fig:selfcorr_turn_metrics}, this setting yields pass rate improvements, with substantial gains from T1 to T2, and continued but comparatively smaller improvements from T2 to T3 for several models (e.g., Claude-Opus-4.6).

F2P keeps improving across rounds, suggesting that agent–model pairs can better leverage the feedback provided by Self-Correction setting to satisfy previously failing requirements and complete the task more effectively.

Meanwhile, P2P stays high overall, but different agent-model pairs diverge at different rounds. 
Codex with GPT-5.3-Codex pushes P2P pass rate to 95\% at T2 and 100\% at T3.
In contrast, GPT-5.1-Codex-Max reaches 85\% at T2 but declines to 80\% at T3, and similarly, Claude-Opus-4.6 reaches 95\% at T2 but drops to 90\% at T3.
This suggests that the last self-correction round can widen the change scope. It may complete the remaining hard cases, but it can also introduce new regression testing risks.

In terms of time overhead, multi-round self-correction shows mixed effects on runtime across different agents and models, with no consistent trend of increasing time cost in later rounds. The gains are real, but the marginal benefit tends to shrink in later rounds.

\subsection{Human-Agent Collaboration}
To investigate the performance of human-agent collaboration, we designed two experimental setups:

\noindent\textbf{Static Plan Injection.} Before the execution phase, we inject key plans without specific code details. This setup evaluates whether reducing planning errors leads to improved task accuracy, thereby quantifying the impact of Planning Capability.

\noindent\textbf{Dynamic Interactive Guidance.} In this setting, the model autonomously decides whether to request human intervention based on its current state. Upon intervention, humans provide the next steps and future roadmap, offering explanatory guidance rather than direct implementation. If the model explicitly asks for code, the request is denied to force the model to reason independently. The maximum number of interventions is limited to 3.

\begin{table}[t]
\centering
\caption{Results of Human-Agent Collaboration. Pass and F2P Score are reported as percentages (\%). Time is measured in minutes. Inter.Avg represents the average number of human interventions per task.}
\label{tab:human_guidance_task_success}
\resizebox{\linewidth}{!}{
\setlength{\tabcolsep}{3pt}
\begin{tabular}{llcccc}
\toprule
Agent \& Model & Mode & Pass & F2P Score & Time & Inter.Avg \\
\midrule
\multirow{5}{*}{\makecell{Codex \\ GPT-5.3-Codex}} & Base (Avg) & 15.0 & 44.1 & 8.8 & - \\
 & Self-Correction & 40.0 & 58.4 & 26.7 & - \\
 & Plan & 41.7 & 59.3 & 10.7 & - \\
 & Interactive & 45.0 & 61.3 & 25.9 & 2.7 \\
 & Plan \& Interactive & 50.0 & 62.3 & 22.4 & 2.2 \\
\midrule
\multirow{5}{*}{\makecell{Claude Code \\ Claude-Opus-4.6}} & Base (Avg) & 16.7 & 50.7 & 17.6 & - \\
 & Self-Correction & 55.0 & 63.6 & 43.8 & - \\
 & Plan & 58.3 & 65.4 & 15.4 & - \\
 & Interactive & 58.3 & 67.4 & 39.6 & 2.4 \\
 & Plan \& Interactive & 61.7 & 69.3 & 35.4 & 2.1 \\
\bottomrule
\end{tabular}
}
\end{table}

Table~\ref{tab:human_guidance_task_success} summarizes performance across two agent-model pairs (Codex with GPT-5.3-Codex; Claude Code with Claude-Opus-4.6), showing that current agents face dual limitations in both planning and execution proficiency, and human collaboration significantly mitigates these bottlenecks.

Plan injection outperformed the self-correction baseline in both pass rate and efficiency.
For instance, Claude Code with plan injection achieved a 58.3\% pass rate compared to 55.0\% with self-correction.
This confirms that establishing a correct plan upfront is far more efficient than relying on the agent to iteratively fix errors during the execution process, as it prevents the agent from entering incorrect logical branches initially.

Interactive guidance generally achieved higher performance compared to static plan injection alone. 
Claude Code improved from 58.3\% (Plan) to 58.3\% (Interactive) in pass rate, while F2P Score increased from 65.4 to 67.4.
While static plans provide a roadmap, they cannot anticipate all runtime anomalies. 
This dynamic intervention proves more robust, as it effectively guides the agent away from unproductive execution trajectories and logic errors that a static plan cannot anticipate.

The combined Plan \& Interactive setting yielded the best results across all metrics.
Crucially, the presence of a pre-injected plan reduced the need for human intervention (e.g., Inter.Avg dropped from 2.4 to 2.1 for Claude Code), indicating a synergy where the plan handles the roadmap while human interaction resolves specific execution problems.

Our findings indicate that rather than exclusively chasing full autonomy, future work should focus on developing collaborative systems that leverage the synergy between efficient execution and human strategic guidance.

\subsection{Error Analysis}
To better understand why current agents achieve low end-to-end success on LongCLI-Bench (most pass rates <20\%), we manually inspected 50 failed trajectories across representative agent-model pairs (primarily Codex and Claude Code). Overall, \textbf{failures are rarely caused by local syntax-level coding mistakes.} Instead, they are dominated by long-horizon workflow breakdowns, where the agent must plan, verify state, and preserve consistency across many interdependent steps. This aligns with our step-level findings that the majority of runs remain below 30\% completion.

Overall, LongCLI-Bench failures are primarily caused by the following three reasons:

\noindent\textbf{Repetitive loops from weak strategic adaptation.} A common pattern is that the agent encounters an execution failure, proposes a superficial patch, re-runs the same command, observes the same error, and repeats until the step limit is exhausted. Such trajectories reveal that the agent fails to recognize that the current plan is invalid and does not shift focus to addressing the root cause.

\noindent\textbf{Environment grounding and verification gaps.} We observed instances of misdiagnosis in which environment-related issues were incorrectly attributed to code logic flaws, resulting in code edits that failed to address the actual cause of the failure.

\noindent\textbf{Long-term inconsistency and regression.} Even when agents make substantial progress on the new requirements, they often break existing functionality: while the average P2P step scores are very high, the P2P pass rates are much lower, indicating that agents introduce regressions risks when they make non-trivial edits. We also saw context drift in long runs (e.g., forgetting earlier constraints), which explains why later self-correction rounds can widen change scope and increase regression risks.

\section{Conclusion}

LongCLI-Bench introduces a new benchmark for evaluating agentic programming in long-horizon command-line interface (CLI) tasks. By curating 20 complex tasks from over 1,000 real-world workflows and computer science assignments, LongCLI-Bench addresses key gaps in existing benchmarks, such as their focus on short, isolated tasks. Unlike many benchmarks that offer binary pass/fail evaluations, LongCLI-Bench emphasizes long-duration tasks that reflect real-world challenges and evaluates both requirement fulfillment and regression avoidance through step-level scores.
The results show that current agent systems struggle with long-horizon tasks, achieving low pass rates. However, human-agent collaboration significantly boosts performance, highlighting the need to prioritize collaborative workflows in future research alongside autonomous agent advancements.

\section*{Limitations}
\noindent\textbf{Task Creation Requires Significant Manual Effort:} Creating tasks for LongCLI-Bench involves extensive manual work, including writing requirement documents, constructing solution paths, setting up test environments, and creating test scripts. This process is time-consuming, with each task taking an average of 40 hours to complete, resulting in a relatively small dataset. Additionally, this makes it challenging to curate high-quality, long-horizon tasks that effectively challenge agents.

\noindent\textbf{Evaluation Metrics:} While step-level scores provide more granular insights, they do not fully assess agent performance in areas like code quality or efficiency. Future versions of the benchmark could include additional metrics for a more complete evaluation.

\section*{Ethics Statement}
\noindent\textbf{Data Privacy and Usage: }All tasks in LongCLI-Bench are derived from publicly available resources or manually curated data. No sensitive or proprietary information was used, and all data have been anonymized to ensure privacy.

\noindent\textbf{Human-Agent Collaboration: }This study involves human-agent collaboration to explore how agent performance improves with human guidance. We strictly adhere to ethical guidelines to ensure participant rights and safeguard their privacy.

\noindent\textbf{Evaluation Ethics: }All data used for evaluation are standardized and anonymized, ensuring compliance with ethical research practices.

\noindent\textbf{Transparency: }We will release detailed task creation processes and evaluation results to ensure transparency and encourage open participation in AI research.

\bibliography{custom}

\end{document}